\documentclass[a4paper,twocolumn, accepted=2025-02-17]{quantumarticle}
\pdfoutput=1
\PassOptionsToPackage{compress}{natbib}
\usepackage[numbers]{natbib}
\usepackage[latin9]{inputenc}
\usepackage{graphicx}
\usepackage{amsmath}
\usepackage{amssymb}
\usepackage{verbatim}
\usepackage{enumerate}
\usepackage{mathtools}
\usepackage{amsthm}
\usepackage{hyperref}
\usepackage{braket}
\usepackage{cancel}
\usepackage{dsfont}
\usepackage{color}

\newtheorem{theorem}{Theorem}

\newtheorem{definition}{Definition}
\newtheorem*{theorem1}{Theorem 1}
\newtheorem*{theorem2}{Theorem 2}

\newcommand{\id}{\mathds{1}}  
\newcommand{\Tr}{\mathrm{Tr}}
\newcommand{\ie}{\textit{i.e. }}
\newcommand{\eg}{\textit{e.g. }}

\begin{document}

\title{On the locality of qubit encodings of local fermionic modes}
\date{17 February 2025}
\author{Tommaso Guaita}
\email{tommaso.guaita@fu-berlin.de}
\affiliation{Dahlem Center for Complex Quantum Systems, Freie Universit\"{a}t Berlin, Germany}
\orcid{0000-0001-9367-1249}

\begin{abstract}
Known mappings that encode fermionic modes into a bosonic qubit system are non-local transformations. In this paper we establish that this must necessarily
be the case, if the locality graph is complex enough (for example for regular 2$d$ lattices). In particular we show that, in case of exact encodings, a fully local mapping is possible
if and only if the locality graph is a tree. If instead we allow ourselves to also consider operators that only act fermionically on a subspace of the qubit Hilbert space, then we show
that this subspace must be composed of long range entangled states, if the locality graph contains at least two overlapping cycles. This implies, for instance, that on 2$d$ lattices
there exist states that are simple from the fermionic point of view, while in any encoding require a circuit of depth at least proportional to the system size to be prepared.
\end{abstract}

\maketitle

\section{Introduction}

Fermi statistics are one of the two fundamental types of particle statistics. Despite their fundamental nature, however, fermionic particles are often pictured as intrinsically
non-local objects. Indeed, creation and annihilation operators of fermionic modes obey canonical anticommutation relations, which are at odds with standard definitions
of locality. This tension can be resolved by postulating that the only \emph{physical} fermionic operators are the ones composed of products of an even number of creation
or annihilation operators. These operators always commute if they are associated to spatially separated modes, giving rise to a system that can be interpreted as a fully 
local model of quantum computation~\cite{bravyi_fermionic_2002, lloyd_unconventional_2000}. 

Nonetheless, the intuition of fermions as non-local particles re-emerges when one tries to simulate this model of computation with a model based on local qubits (or in general
on local finite-dimensional Hilbert spaces). The most basic way to do this is the Jordan-Wigner transformation~\cite{jordan_uber_1928}, which for sufficiently non-trivial systems, such as
2$d$ lattices, will always encode some local fermionic operators into qubit operators with a large non-local support (in fact a support growing with the system's size).
More sophisticated encodings exist which are able to circumvent this problem by introducing ancillary qubits (\eg the Bravyi-Kitaev superfast encoding~\cite{bravyi_fermionic_2002},
the auxiliary fermion encoding~\cite{verstraete_mapping_2005,ball_fermions_2005} or other similar constructions~\cite{wosiek_local_1982, szczerba_spins_1985, setia_superfast_2019,
whitfield_local_2016, havlicek_operator_2017, chen_exact_2018, bochniak_constraints_2020, derby_compact_2021, li_higher-dimensional_2022,
chien_optimizing_2022, chen_equivalence_2023, nys_quantum_2023}): here the total dimension of the qubit Hilbert space is larger that the one of the fermionic space, but it is   
possible to define some local qubit operators that implement the fermionic algebra on a subspace of qubit states, which we intrepret as representing the
physical fermionic states. While this resolves the locality issue at the operator level (all local fermionic operators are mapped to local qubit operators), the same
issue reappears now at the level of the encoded states. The subspace that encodes fermionic states is indeed in general composed of non-local qubit states, with correlations extending across
the whole system.    

These encodings are of great value as they prove the equivalence of the models of computation based on fermions and qubits, up to an overhead scaling polynomially
with the system size. But, at the same time, they also highlight how this equivalence necessarily involves some amount of non-locality. The main aim of this paper is to 
provide a systematic understanding of this non-locality within a simple framework, showing when and how it emerges and providing bounds on the minimal amount of it that is
necessary to encode fermionic modes correctly.

More precisely, we consider systems of an arbitrary fixed geometry, represented by a graph (for instance a 2$d$ lattice or a 1$d$ chain), and we try to encode some local fermionic modes
of this graph into qubit operators that are as local as possible with respect to the \emph{same} geometry. Maintaining the same locality graph for both the fermionic and qubit system
allows a fair comparison of the resources, in terms of locality, that are necessary to implement both cases. This approach allows us to include in our discussion most existing encodings. 
It is however worthwhile to note that there exist also encodings where the
qubit operators live in a system with a completely different locality structure compared to the fermions that they represent. In the most common of these, the qubits are distributed
on a Fenwick tree structure (initally introduced by Bravyi and Kitaev~\cite{bravyi_fermionic_2002} and further elaborated in Refs.~\cite{havlicek_operator_2017,jiang_optimal_2020}). 
While the latter approach may be very practical in certain applications, it however implies that the local structure of the fermions is violated already from the onset when constructing
these encodings.
Notice further that in these constructions local fermionic operators are in general still mapped to qubit operators with a support growing with the systems size, albeit only logarithmically.
We will not analyse this case further. 

In what follows, we provide instead a framework to define encodings for a fixed geometry and assess all the sources of non-locality that appear in their construction.
We will prove two main results which establish that the minumum amount of non-locality that an encoding needs is related to simple properties of the system's geometry graph.
In particular, we will show that a fully local encoding is possible only for tree graphs. If the graph contains cycles, the best that can be achieved is to ensure that local fermionic operators
are mapped to local qubit operators. This however may come at the price of encoded states containing non-local correlations. The minimal extent of these correlations is again related
to certain properties of the graph, which we identify. If the graph contains at least two cycles that overlap, then no encoded state can be a product state. Rather, all encoded states have to be constructed
by applying a circuit of local gates to a product state, the minimal depth of which circuit grows linearly with a quantity $d$, representing a certain notion of dimension of the maximal overlap
between any two cycles in the graph.

These results have relevant consequences both at the fundamental and the practical level. At the fundamental level they give us a new insight on the deep distinction between fermionic
and bosonic systems. While the former may arise as excitations of the latter, this can happen only after sufficiently non-local correlations have spread across the system. At a more
practical level, these results provide some bounds on the possibility of simulating fermions using qubit-based quantum computing platforms rather than natively fermionic ones. 
We show that this necessarily requires linear overheads which can be problematic in near term devices, where noise can quickly propagate across circuits of very modest depth~\cite{stilck_franca_limitations_2021, de_palma_limitations_2023}.
  
The rest of the paper is structured as follows. In Section~\ref{sec:def}, we introduce all the necessary notation and define precisely what we mean by an encoding of local fermionic modes into
a qubit Hilbert space. In Section~\ref{sec:results}, two theorems are presented which constitute the main results of the paper. In Section~\ref{sec:discussion} we discuss and summarise the 
consequences of these results. In Appendix~\ref{sec:proofs} we provide the proofs of the theorems stated above.

\section{Notation and Definitions} \label{sec:def}
A system of $N$ fermionic modes is abstractly defined by the creation annihilation operators $a_k^\dag$, $a_k$ for each mode $k=1,\dots,N$. We can endow this system
with a notion of locality by considering an underlying graph $(V,E)$, where $V$ is the set of vertices and $E\subset V\times V$ the set of edges of the graph. If $(j,k)\in E$ 
for some vertices $j,k\in V$, it means that these two vertices are ``neighbouring'' according to the locality defined by the graph. We assume that the graph has exactly $N$ 
vertices and that each fermionic mode is associated to one of these vertices. In this case we talk about a system of \emph{local fermionic modes}~\cite{bravyi_fermionic_2002}.
The space of states on which these mode operators act is conventionally called the fermionic Fock space $\mathcal{F}$. It can be constructed by considering a vacuum state
$\ket{0}$ (satisfying $a_k\ket{0}=0$ for all $k$) and all other states obtained by acting on $\ket{0}$ with all possible combinations of $a_k^\dag$ and $a_k$. It can be shown that this space has dimension $2^N$ (see for example Ref.~\cite{vidal_quantum_2021} for more details).

The Fock space $\mathcal{F}$ can be equivalently described in terms of the Majorana operators
\begin{align}
    c_k &= (a_k^\dag + a_k), \\
    c_{N+k} &= i(a_k^\dag - a_k) \quad \mbox{ for } k=1,\dots,N.
\end{align}
The Majorana operators are Hermitian and satisfy the canonical anticommutation relations $\{c_i,c_j\}=2\delta_{ij}$, which fully specify the structure of the group of even Majorana monomials:
\begin{align}
    \mathcal{M}_+ &=\left\{ \, \alpha \prod_{i=1}^{2N} c_i^{n_i} \, \Big| \, \alpha\in\left\{\pm,\pm i \right\}, \; n\in \{0,1\}^{2N}\!\!, \right. \nonumber\\
    &\left. \hspace{29mm} \mbox{s.t. } \sum_i n_i \; \mathrm{mod} \; 2 =0 \right\} .
\end{align}
One of the main objects of interest is the algebra of \emph{fermionic observables}, which is defined as all complex linear combinations of even Majorana monomials
in $\mathcal{M}_+$.

The fermionic observables are endowed with a local structure, in the sense that they can be generated by a set of local objects associated with the local structures
of the underlying graph. Indeed,  if the fermionic modes are associated to a \emph{connected} graph $(V,E)$, then, following the notation of Ref.~\cite{bravyi_fermionic_2002}, a set of generators of the group $\mathcal{M}_+$ is given by the operators
\begin{align}
    A_{jk} &= - A_{kj} = -i c_j c_k  \hspace{0.1cm}\mbox{for every edge } (j,k)\in E, \label{eq:def-A}\\
    B_k &= -i c_k c_{N+k}            \hspace{0.9cm}\mbox{for every vertex } k\in V.
\end{align}
They are Hermitian and satisfy the following relations, which fully specify the group structure 
\begin{align}
    [B_k ,B_j]&=0, \label{eq:group-relations-AB_i}\\
    A_{jk} B_l - (-1)^{\delta_{jl}+\delta_{kl}} B_l A_{jk} &=0,\label{eq:group-relations-AB_AB}\\
    A_{jk} A_{lm} - (-1)^{\delta_{jl}+\delta_{kl} + \delta_{jm}+\delta_{km}} A_{lm} A_{jk} &=0, \label{eq:group-relations-AB_AA} \\
    A_{jk}^2=B_k^2&=\id, \label{eq:group-relations-AB_BB}\\
    i^n A_{j_1,j_2} A_{j_2, j_3} \cdots A_{j_n, j_1} &= \id  \label{eq:group-relations-AB_f}
\end{align}
for every closed path $j_1, j_2, \dots, j_n$ in the graph. Relations~\eqref{eq:group-relations-AB_AB} and~\eqref{eq:group-relations-AB_AA} imply that the operators $A_{jk}$ and $B_k$ anticommute if they are associated to incident edges/vertices and they commute otherwise. The operators $B_k$ always commute among themselves.
We will refer to this set of generators as \emph{local fermionic generators}. Notice that local generators associated to geometrically separated vertices/edges always
commute, compatibly with basic notions of locality\footnote{This is the reason why we chose to only consider \emph{even} Majorana monomials as observables. Generators
with an odd number of Majorana would not commute even if localised far apart, breaking basic locality assumptions. This is consistent with the common `superselection rule' that 
all physically relevant observables and Hamiltonians should be even Majorana polynomials~\cite{vidal_quantum_2021}. Notice therefore that the non-locality of encodings that we 
will discuss below is not just a trivial consequence of having misguidedly included odd Majorana terms in our analysis.}.

Consider now another system described by the Hilbert space $\mathcal{H}$ (this could be, for example, a register of qubits). We want to use the system $\mathcal{H}$ to represent
the fermionic system. In other words we want to \emph{encode} $\mathcal{F}$ in $\mathcal{H}$. Following Refs.~\cite{chien_optimizing_2022, bravyi_fermionic_2002}, an encoding is
an algebra isomorphism between the algebra of fermionic observables and an algebra of operators in $\mathrm{L}(\mathcal{H})$. To obtain this, it is sufficient to have a faithful
representation of the group $\mathcal{M}_+$ on $\mathcal{H}$, which can then be extended to an algebra isomorphism by linearity\footnote{Note that any representation of $\mathcal{M}_+$
has to be faithful. Indeed, for any non-trivial element $g$ of $\mathcal{M}_+$ it is possible to find another group element that anticommutes with it, so the representative of $g$
cannot be the identity. So the faithfulness requirement is actually trivial.}. In other words, it is sufficient to have a
representation of the generators $A_{jk}$ and $B_k$ as operators on $\mathcal{H}$. We indicate the corresponding representatives as $\hat{A}_{jk}$ and $\hat{B}_k$.

\begin{definition}[Fermionic encoding]
Consider the Fock space $\mathcal{F}$ of $N$ local fermionic modes on a connected graph $(V,E)$. An \emph{encoding} of $\mathcal{F}$ in the Hilbert space $\mathcal{H}$
is defined by providing the Hilbert space operators $\hat{B}_k\in \mathrm{L}(\mathcal{H})$ for each vertex $k\in V$ and $\hat{A}_{jk} \in \mathrm{L}(\mathcal{H})$ for each edge $(j,k)\in E$,
such that $\hat{A}_{jk}^\dag=\hat{A}_{jk}$, $\hat{B}_k^\dag = \hat{B}_k$, $\hat{A}_{jk}=-\hat{A}_{kj}$ and the relations~\eqref{eq:group-relations-AB_i}--\eqref{eq:group-relations-AB_f}
are satisfied. \label{def:encoding}
\end{definition}
\noindent Notice that having such an encoding of the operators $\hat{A}_{jk}$ and $\hat{B}_k$ implies that it is possible to construct a subspace of $\mathcal{H}$ isomorphic to the Fock space
$\mathcal{F}$ (or at least its even/odd parity subspace). Any quantum computation performed on this subspace will give the same outcome as one performed on $\mathcal{F}$, as this
outcome is ultimately determined by the group relations satisfied by $\hat{A}_{jk}$ and $\hat{B}_k$, which we assume to be the correct ones.

Consider now the case in which also the Hilbert space $\mathcal{H}$ has a locality structure given by the same graph $(V,E)$ as the fermionic modes.
By this we mean that $\mathcal{H}$ has a tensor product structure
\begin{equation}
    \mathcal{H}=\mathcal{H}_{k_1} \otimes \mathcal{H}_{k_2} \otimes \cdots \otimes \mathcal{H}_{k_N}\,, \label{eq:local-hilbert-space}
\end{equation}
where each $\mathcal{H}_k$ is a local Hilbert space associated to each of the graph's vertices $k\in V$. 
For simplicity, we will assume that $\mathcal{H}_k$ has a finite dimension $d_k<+\infty$. However the following results should hold similarly also for infinite
dimensional local spaces, which arise for instance in systems of indistinguishable bosonic particles.
The structure~\eqref{eq:local-hilbert-space} arises, for example, if we attach a certain number of qubits to each vertex of the graph (or in general any number of finite dimensional systems
-- not just qubits -- can be attached to each vertex, as $d_k$ is arbitrary). This structure directly applies to many existing constructions: for example, the Verstraete-Cirac
encoding~\cite{verstraete_mapping_2005} attaches two qubits to each fermionic vertex in a 2$d$ geometry. Some encodings attach qubits to edges rather than vertices (for example
the original Bravyi-Kitaev superfast encoding~\cite{bravyi_fermionic_2002}), but these can quite simply be reformulated in a way that is compatible with our language (see Ref.~\cite{setia_superfast_2019}), so their analysis is not fundamentally different.

In this setting an operator on $\mathcal{H}$ is local if it has support only on local Hilbert spaces associated to ``neighbouring'' vertices according to the graph's
connectivity. To make this more precise, consider a basis $\left\{ \sigma_{k,a} \right\}_{a=1,\dots,d_k^2}$ of the space $ \mathrm{L}(\mathcal{H}_k)$ of linear operators
on the local Hilbert space $\mathcal{H}_k$, for each vertex $k$. 
The operators $\sigma_{k,a}$ can be chosen to be Hermitian. In the case that $\mathcal{H}_k$ is a single qubit Hilbert space they could be, for example, the Pauli matrices
together with the identity. We can now define the spaces of local operators associated to each vertex $k$ or edge $(j,k)$ respectively as
\begin{align}
    \ell_k(\mathcal{H})&=\mathrm{span} \left\{ \id\otimes  \cdots \otimes \sigma_{k,a} \otimes \cdots \otimes \id \,, \right. \nonumber \\
    &\hspace{33mm} \left. \mbox{for } a=1,\dots,d_k^2 \right\} \\[3mm]
    \ell_{jk}(\mathcal{H})&=\mathrm{span} \left\{ \id\otimes  \!\cdots\! \otimes \sigma_{j,a} \otimes \!\cdots\! \otimes  \sigma_{k,b} \otimes \!\cdots\! \otimes \id\,, \right. \nonumber\\
    &\hspace{17mm} \left. \mbox{for } a=1,\dots,d_j^2, \; b=1,\dots,d_k^2 \right\}.
\end{align}
\noindent A local encoding of $\mathcal{F}$ in $\mathcal{H}$ is an encoding where the representatives $\hat{A}_{jk}$ and $\hat{B}_k$ of local fermionic operators are local also according to this tensor
product structure of $\mathcal{H}$.

\begin{definition}[Local fermionic encoding]
Consider the Fock space $\mathcal{F}$ of $N$ local fermionic modes on a connected graph $(V,E)$ and a Hilbert space $\mathcal{H}$ with locality structure also given by $(V,E)$ according
to~\eqref{eq:local-hilbert-space}. An encoding of $\mathcal{F}$ into $\mathcal{H}$, defined as in Definition~\ref{def:encoding}, is \emph{local} if $\hat{B}_k\in \ell_k(\mathcal{H})$ for
each vertex $k\in V$ and $\hat{A}_{jk} \in \ell_{jk}(\mathcal{H})$ for each edge $(j,k)\in E$. \label{def:local-encoding}
\end{definition}

In some situations, it may be sufficient to consider local operators $\hat{A}_{jk}$ and $\hat{B}_k$ that generate a representation of $\mathcal{M}_+$ only restricted to a subspace $\mathcal{C}$
of $\mathcal{H}$. Only states in this subspace would then represent physical fermionic states. This would be relevant, for example, in the case of digital quantum computation if it is possible
to generate initial states in $\mathcal{C}$, which would then remain in $\mathcal{C}$ throughout the computation. It would also be relevant in the case of analog quantum simulation if it is
possible to engineer the system Hamiltonian to restrict evolutions within $\mathcal{C}$, for example by adding terms that penalise in energy states outside $\mathcal{C}$. Let us call this
type of encoding a \emph{local block encoding}.

\begin{definition}[Local fermionic block encoding] \label{def:gen-local-encoding}
Consider the Fock space $\mathcal{F}$ of $N$ local fermionic modes on a connected graph $(V,E)$ and a Hilbert space $\mathcal{H}$ with locality structure also given by $(V,E)$ according
to~\eqref{eq:local-hilbert-space}. A \emph{local block encoding} of $\mathcal{F}$ into $\mathcal{H}$ is defined by providing the Hilbert space operators $\hat{B}_k\in\ell_k(\mathcal{H})$
for each vertex $k\in V$ and $\hat{A}_{jk} \in \ell_{jk}(\mathcal{H})$ for each edge $(j,k)\in E$ such that there exists a subspace $\mathcal{C}$, closed under the actions of
$\hat{A}_{jk}$ and $\hat{B}_k$, where for every $\ket{\psi}\in\mathcal{C}$
\begin{align}
    [\hat{B}_k ,\hat{B}_j]\ket{\psi}&=0 \label{eq:gen-relations-AB_i}\\
    \left[\hat{A}_{jk} \hat{B}_l - (-1)^{\delta_{jl}+\delta_{kl}} \hat{B}_l \hat{A}_{jk} \right] \ket{\psi} &=0 \label{eq:gen-relations-AB_AB}\\
    \left[\hat{A}_{jk} \hat{A}_{lm}  \!-\! (\!-1)^{\delta_{jl}+\delta_{kl} + \delta_{jm}+\delta_{km}} \!\hat{A}_{lm} \hat{A}_{jk} \!\right] \! \ket{\psi} \!&=0 \label{eq:gen-relations-AB_AA} \\[1mm]
    (\hat{A}_{jk}^\dag-\hat{A}_{jk})\ket{\psi}=(\hat{B}_k^\dag-\hat{B}_k)\ket{\psi}&=0 \label{eq:gen-relations-AB_h}\\[2mm]
    \hat{A}_{jk}^2\ket{\psi}=\hat{B}_k^2\ket{\psi}&=\!\!\ket{\psi} \label{eq:gen-relations-AB_BB}\\[2mm]
    i^n \hat{A}_{j_1,j_2} \hat{A}_{j_2, j_3} \cdots \hat{A}_{j_n, j_1} \ket{\psi} &= \!\! \ket{\psi}  \label{eq:gen-relations-AB_f}
\end{align}
for every closed path $j_1, j_2, \dots, j_n$ in the graph.This essentially means that only the subblock of the local operators $\hat{A}_{jk}$ and $\hat{B}_k$ corresponding to the subspace $\mathcal{C}$ acts as a fermionic encoding.
\end{definition}

Let us conclude this section by stressing that we have introduced here two distinct notions of locality. In the Fock space setting, we consider operators such as $A_{jk}$ and $B_k$ to be local
because they can be associated to local objects (vertices, edges etc.) and they commute if the corresponding local objects are geometrically separated. Local operators of this type arise
naturally in systems that contain physical fermionic particles. In the Hilbert space setting, we have defined a stronger notion of locality which requires the Hilbert space to decompose
into a tensor product structure. This structure arises naturally in systems built by assembling several local subsystems (\textit{e.g.} qubits).
Notice that the latter definition of locality implies the former. On the other hand, we will show below that the converse is not true, that is local systems in 
the first sense do not always admit the structure of the second.

\section{Results}\label{sec:results}
In the previous section we have introduced the definitions of some relevant types of fermionic encodings. We will now discuss some results about the feasibility and complexity of constructing and
implementing these types of encodings.

First of all, if one talks only generically of encodings, such as the ones of Definition~\ref{def:encoding}, with no further constraints, then it is well established that the fermionic
Fock space can be encoded into other non-fermionic Hilbert spaces. For instance, the well-known Jordan-Wigner transformation encodes $\mathcal{F}$ into the Hilbert space
of $N$ qubits, for any locality structure that one may wish to impose on the fermionic modes. For simple geometries, such as, for example, if the graph $(V,E)$ represents a chain
with open boundaries, then it turns out that the Jordan-Wigner transformation is even a local encoding, according to Definition~\ref{def:local-encoding}.

However, as soon as one considers slightly less trivial geometries (\eg closed rings), it is clear that the Jordan-Wigner encoding becomes highly non-local, with some local fermionic
operators $A_{jk}$ being mapped to non-local operators with support potentially on the whole system. In fact, for non-trivial enough geometries it is impossible to encode fermions locally.
This turns out to be related to the presence of closed loops in the locality graph:

\begin{theorem}[No local encodings for cyclic graphs]
A set of local fermionic modes admits a local encoding according to the Definitions~\ref{def:local-encoding} if and only if their locality graph $(V,E)$ is a tree graph.
A tree graph is a connected graph that contains no cycles, that is no sequences of edges $\{(j_1,j_2), (j_2,j_3),\dots,(j_p,j_1)\}$ starting and ending in the same vertex
without any edge appearing more than once. \label{thm:no-local-encoding}
\end{theorem}

This theorem formalises the fact, already stated above, that the notion of locality inherent to local fermionic modes is strictly weaker than the one coming from a tensor product Hilbert space
as in~\eqref{eq:local-hilbert-space}. In other words, the result gives a more precise formulation of the statement that ``fermionic Fock space does not admit a tensor product structure'',
which is sometimes encountered. A proof of the theorem is given in Appendix~\ref{sec:proof-locality}: it essentially relies on the fact that it is always possible to construct locally anti-commuting
operators, however such operators can never satisfy condition~\eqref{eq:group-relations-AB_f}, if they are local in the tensor product sense.

Next to the Jordan-Wigner transformation, there exist in the literature several further attempts at defining local encodings of fermionic modes, the most famous of which are the Bravyi-Kitaev superfast encoding~\cite{bravyi_fermionic_2002, setia_superfast_2019}
or the Verstraete-Cirac encoding~\cite{verstraete_mapping_2005}. These encodings all avoid the constraints represented by Theorem~\ref{thm:no-local-encoding} by not constructing a local encoding
according to Definition~\ref{def:local-encoding} but rather a local \emph{block} encoding, as introduced in Definition~\ref{def:gen-local-encoding}. We recall that this means that the local
operators $\hat{A}_{jk}$ and $\hat{B}_k$ introduced in these encodings do not satisfy the relations~\eqref{eq:group-relations-AB_i}--\eqref{eq:group-relations-AB_f} on the full Hilbert space
$\mathcal{H}$, but rather only on a specific subspace $\mathcal{C}$. It follows from Theorem~\ref{thm:no-local-encoding} however that this subspace $\mathcal{C}$ cannot have a
local structure, in particular it cannot admit a tensor product decomposition into local subspaces. So it may seem that these block encodings are effectively just shifting the non-locality
of the encoding from the operators $\hat{A}_{jk}$ and $\hat{B}_k$ to the states in the subspace $\mathcal{C}$. To verify to what extent this is the case, it would be helpful to understand how
non-local exactly these states in $\mathcal{C}$ are. In what follows we will introduce a result that quantifies this.

It is again the case that the non-locality of the encoding ultimately depends on the
geometry of the locality graph $(V,E)$. However, unlike the case of the previous theorem, the complexity here is not directly related to the presence of individual cycles in
the graph. Indeed, for graphs containing a single loop, such as a ring graph,
it is possible to define a block encoding where $\mathcal{C}$ contains sufficiently local states: for instance, where the
fermionic vacuum is encoded into a product state with respect to the local structure~\eqref{eq:local-hilbert-space} of $\mathcal{H}$ (see Appendix~\ref{sec:one-loop} for an explicit construction of this). But the states in $\mathcal{C}$ do actually start to 
become progressively more distant from product states as soon as the graph contains at least two overlapping cycles. The larger these overlapping cycles are, the deeper the circuits will be which
are needed to produce any state in $\mathcal{C}$ out of a product state. This concept of overlapping loops, as well as the one of their size, are made more precise in the following definitions.

\begin{figure}
\centering
\includegraphics[width=7cm]{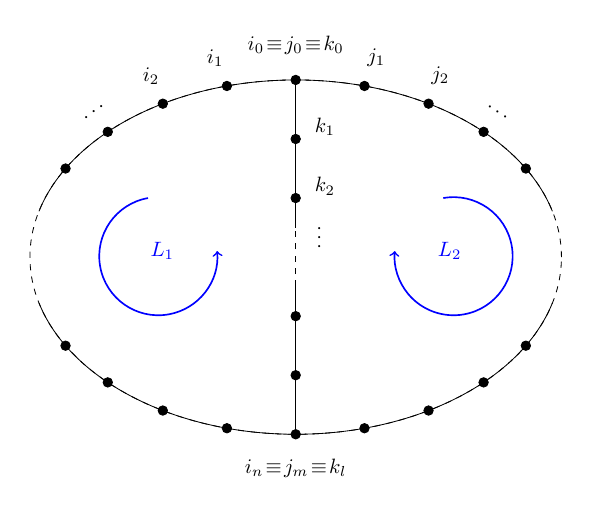}
\caption{Structure of an 8-shaped graph. The sequences $\{i_0, i_2, \dots, i_n\}$, $\{j_0, j_2, \dots, j_m\}$ and $\{k_0, k_2, \dots, k_l\}$ share an initial point $i_0\equiv j_0\equiv k_0$ and
          a final point $i_n\equiv j_m\equiv k_l$. This graph contains two independent overlapping cycles with base point $i_0$: $L_1$ follows first the $i$ vertices then the $k$ vertices; $L_2$ 
          first the $j$ vertices and then the $k$ vertices.}
\label{fig:figure-8}
\end{figure}

\begin{definition}[8-shaped subgraph] \label{def:8-shaped-subgraph}
    Consider a connected graph $(V,E)$. If it is possible to identify in the graph three disjoint sequences of edges
    \begin{align}
        &\{(i_0,i_1), (i_1,i_2),\dots,(i_{n-1},i_n)\}\,, \\
        &\{(j_0,j_1), (j_1,j_2),\dots,(j_{m-1},j_m)\}\,, \\
        &\{(k_0,k_1), (k_1,k_2),\dots,(k_{l-1},k_l)\}
    \end{align}
    such that $i_0\equiv j_0\equiv k_0$ and $i_n\equiv j_m\equiv k_l$ then we say the graph has an \emph{8-shaped} subgraph. This means that it is possible to find a subgraph with the structure shown
    in Figure~\ref{fig:figure-8}, \ie containing two overlapping loops. For any integer $D<n,m,l$ we indicate the sets of the first $D$ edges of each sequence as
    \begin{align}
        I_D&=\{(i_0,i_1), (i_1,i_2),\dots,(i_{D-1},i_D)\}\,, \label{eq:ID}\\
        J_D&=\{(j_0,j_1), (j_1,j_2),\dots,(j_{D-1},j_D)\}\,, \\
        K_D&=\{(k_0,k_1), (k_1,k_2),\dots,(k_{D-1},k_D)\} \,.\label{eq:KD}
    \end{align}
    The set of the remaining edges of the subgraph not included in $I_D$, $J_D$ or $K_D$ we indicate as
    \begin{align}
        R_D&=\{(i_{D},i_{D+1}),\dots,(i_{n-1},i_n), \nonumber\\
        &\hspace{10mm} (j_{D},j_{D+1}),\dots,(j_{m-1},j_m),\nonumber\\
        &\hspace{15mm} (k_{D},k_{D+1}),\dots,(k_{l-1},k_l)\} \,. \label{eq:RD}
    \end{align}
\end{definition}

\begin{figure*}
\centering
\includegraphics[width=15cm]{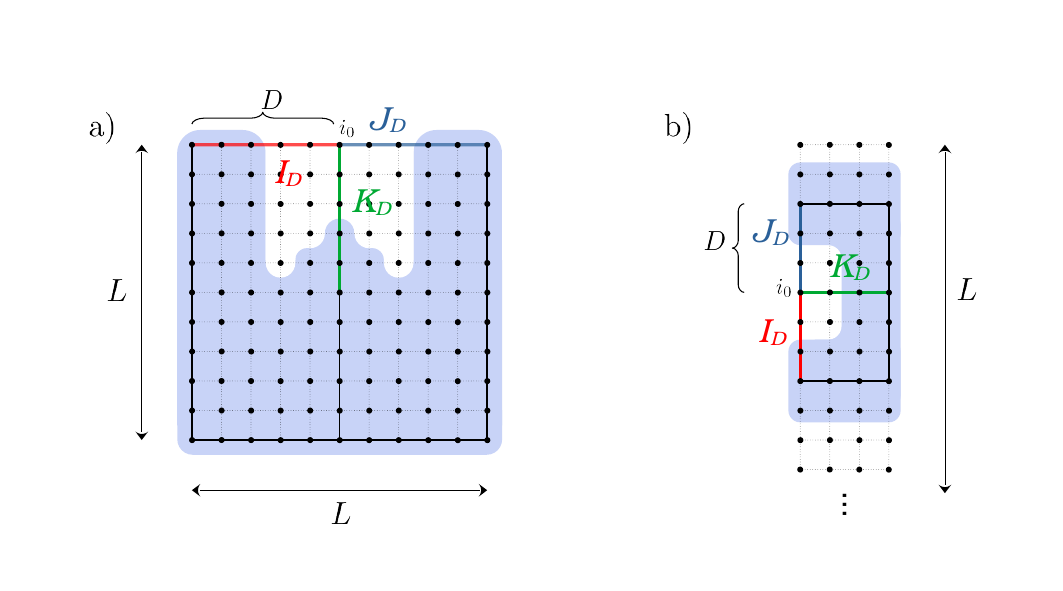}
\caption{Examples of 8-shaped subgraphs within two different graphs: in panel a) an $L\times L$ square lattice, in panel b) a ladder-type graph (that is a square lattice with a fixed width of
 $3$ edges and an arbitrary length $L$). In both graphs the 8-shaped subgraph of largest size is depicted by a black line, it is of size $d=2$ in example a) and size $d=1$ in example b). In both
 graphs a choice of $I_D$, $J_D$ and $K_D$ that achieves this maximal $d$ is indicated in red, blue and green colour. The shaded region represents the vertices that are within distance $d$ of
 $R_D$, which can be seen to all be within distance $d$ of at most one of $I_D$, $J_D$ or $K_D$. In example a) the maximal 8-shaped subgraph has size that grows with $L$, while in example b)
 the represented subgraph of size $d=1$ is maximal for any $L\geq 6$. }
\label{fig:lattice-samples}
\end{figure*}

\begin{definition}[Size of 8-shaped subgraphs] \label{def:8-shaped-size}
    Given an 8-shaped subgraph embedded in a graph $(V,E)$, we say that it has size $d$ if $d$ is the largest integer for which it is possible to find $D\in\mathbb{N}$ such that
    no vertex in the whole graph $V$ is simultaneously within distance $d$ of $R_D$ and within distance $d$ of more than one of $I_D$, $J_D$ or $K_D$. Here distance is to be understood as the graph
    distance, that is the minimum number of edges that are necessary to connect two given vertices.
\end{definition}
This definition of size of the \emph{8-shaped} subgraphs may seem a bit contrived, but it measures in a sense the degree of bidimensionality of the given subgraph. Indeed to maximise
this measure it is necessary for the three sequences that compose the subgraph to be long and to be `far from each other'. So the graph needs to extend into at least two dimensions to allow these
sequences to grow both long and apart from each other at the same time. For example, if the graph is a two-dimensional square lattice of side $L$, see Figure~\ref{fig:lattice-samples}(a),
then the largest 8-shaped subgraph has size $d=\lfloor (L-2)/4 \rfloor$ which grows with the linear size of the system. In general, also for most other truly two-dimensional lattices it is
possible to find 8-shaped subgraphs of size growing with the linear extent of the system. On the other hand, in a ladder-type graph like the one represented
in Figure~\ref{fig:lattice-samples}(b), any 8-shaped subgraph will always have size upper bounded by a constant,  reflecting the fundamentally one-dimensional nature of the graph. 

In any case, the size of the maximal 8-shaped subgraph that a graph can accommodate is directly related to the complexity of the states of any local block encoding of the fermionic modes
on the graph. This is specified by the following theorem:

\begin{theorem}[Complexity of local block encodings]
    Consider the local fermionic modes on a connected graph $(V,E)$ containing an 8-shaped subgraph of size $d$. Consider any local block encoding
    of these modes into a Hilbert space $\mathcal{H}$ according to Definition~\ref{def:gen-local-encoding}. Then the subspace $\mathcal{C}$ does not contain any state that can be produced
    by acting on a product state with a circuit of local two-body gates of depth lower or equal to $d$. \label{thm:encoding-complexity}
\end{theorem}

Here, the product states and the local gates should be understood as defined with respect to the local tensor product structure~\eqref{eq:local-hilbert-space} of the Hilbert
space $\mathcal{H}$. A proof of the theorem is presented in Appendix~\ref{sec:proof-complexity}. The intuitive idea is that if the encoded states are too local, then at the
bifurcation point $i_0\equiv j_0\equiv k_0$ of the three cycles of the 8-shaped subgraph there exist three operators that all anticommute and have a shared localised eigenstate, 
which is impossible. There must exist correlations with the other analogous point $i_n\equiv j_m\equiv k_l$ such that this does not have to be the case. This theorem clarifies 
where the non-locality of local fermionic block encodings lies and what graph structures it is related to.

The results that have been derived so far in this paper can be summarised as in the Table~\ref{table} below. If the graph $(V,E)$ contains no cycles, then there always exist a local encoding
(according to Definition~\ref{def:local-encoding}) of the corresponding fermionic modes. If the graph instead contains cycles, but these do not overlap, then it is possible
to construct local block encodings (according to Definition~\ref{def:gen-local-encoding}), where the vacuum state is however still mapped to a local product state (as discussed
in Appendix~\ref{sec:one-loop}). Finally, if overlapping loops are present in the graph, then in any local block encoding any encoded state will have correlations on a length
scale determined by the size of the largest overlapping loops, as made precise in Theorem~\ref{thm:encoding-complexity}.

\section{Discussion} \label{sec:discussion}

It is indisputable that fermionic observables can be considered as fully local objects, following their own consistent notion of locality. However,
an element of non-locality necessarily emerges when one tries to represent them in a natively bosonic system. The results above prove that non-locality is an essentially inevitable feature of this
type of mappings: it must appear to bridge between the fundamentally incompatible notions of fermionic locality (based on commutation relations) and bosonic locality (based on tensor
products of local Hilbert spaces).

\begin{table*}[th]
\centering
\begin{tabular}{||c | c | c | c||} 
 \hline
  & Local & Local block & Vacuum  \\
 Graph type & encoding & encoding & state  \\[0.8ex] 
 \hline\hline
 No cycles & Yes & Yes & Product \\ 
 Separate cycles & No & Yes & Product \\
 Overlapping cycles & No & Yes & Entangled \\ [1ex] 
 \hline
\end{tabular}
\caption{Summary of the encodings that can be constructed for different types of graphs: tree graphs with no cycles, graphs with individual non-overlapping cycles
and graphs with at least two overlapping cycles.}
\label{table}
\end{table*}

A consequence of the results of this paper is that, in certain geometries, there exist simple fermionic states (for instance that can be created out of the vacuum with a constant
depth circuit of fermionic gates) which in any encoding require a circuit of $\Omega(N)$ depth to be represented. Consider, for instance, a geometry like the one represented in 
Figure~\ref{fig:figure-8}, where each branch of the graph is of equal length. On this system consider the state
\begin{equation}
    \ket{\phi}=\prod_{(j,k)\in E} U_{(j,k)} \ket{0} \mbox{ with } U_{(j,k)}=\exp(i \alpha A_{j,k}),  \label{eq:example-phi}
\end{equation}
for a real constant $\alpha$, where $\ket{0}$ is the fermionic vacuum state. The ordering of the product of non-commuting fermionic gates $U_{(j,k)}$ is chosen such that they form a
circuit of constant depth (for this graph an ordering exists with circuit depth $3$).
So $\ket{\phi}$ can be considered a very simple state from the fermionic point of view. However, if we now look at the states that it would get mapped to by a fermion to boson encoding,
we find the following. If we assume that the encoding preserves the locality of the operators, then by Theorem~\ref{thm:encoding-complexity} any fermionic state, including $\ket{\phi}$,
needs a circuit of depth proportional to $N$ to be produced out of a product state. If, on the other hand, we enforce that $\ket{0}$ is mapped to a product state, giving up if necessary
the locality of the $\hat{A}_{j,k}$ operators, then by a reasoning analogous to the proof of Theorem~\ref{thm:encoding-complexity} it is clear that at least some of the operators $\hat{A}_{j,k}$
must have support of size proportional to $N$. Thus the circuit~\eqref{eq:example-phi} producing $\ket{\phi}$ from the product vacuum state will have, in terms of local 
gates, a depth proportional to $N$.

In the analysis above we have only considered the case in which one constructs the encoded fermionic states out of product states using unitary quantum circuits. It is possible that
the conclusions would change if one also allows measurements and adaptive applications of gates based on the outcomes of these measurements. It is indeed known that long range
entanglement can be created by a finite depth circuit augmented by local measurements and feedforward~\cite{tantivasadakarn_long-range_2022}. This type of protocols have been shown
to efficiently prepare long range correlated states such as the GHZ state~\cite{briegel_persistent_2001, meignant_distributing_2019} or topologically ordered states~\cite{raussendorf_long-range_2005,aguado_creation_2008,piroli_quantum_2021}.
It should be straightforward to extend them to efficiently prepare the encoded states of a local fermionic block encoding.
If this is possible, it would mean that encoded fermionic states are non-trivial with respect to local unitary circuits, but actually live in the trivial phase with respect to local circuits
plus local operations and classical communications (LOCC)~\cite{piroli_quantum_2021}. As a sketch of how this could be done, consider for instance that 
it is always possible to construct encodings that satisfy relations~(\ref{eq:gen-relations-AB_i}--\ref{eq:gen-relations-AB_BB}) with respect to some local operators $\hat{A}_{jk}$
and $\hat{B}_k$ and a product state $\ket{\psi}$. On suitable graphs, it may be further possible to measure all operators $i^n \hat{A}_{j_1,j_2} \hat{A}_{j_2, j_3} \cdots \hat{A}_{j_n, j_1}$
for all cycles of the graph using only constant depth circuits and local measurements, thus projecting $\ket{\psi}$ onto a state that also satisfies relations~\eqref{eq:gen-relations-AB_f} up to
some signs given by the measurement outcomes. Then it would be sufficient to find a way to correct these signs by appropriately reassigning some encoded operators $\hat{A}_{jk} \rightarrow -\hat{A}_{jk}$.

An open question raised by the observations above is whether the fundamental difference between fermions and bosons can be tested or verified in an experimental setting, at least in principle.
If we build a fermionic state through some local actions, and we somehow verify its fermionic nature, can we claim that it must be composed of genuine fermions, and not fermionic
excitations of some ultimately bosonic system? This should be the case because creating fermionic excitations out of a bosonic medium would require some non-local operations which
we have in principle excluded in our preparation. 

Another open question is whether one can build on examples like state~\eqref{eq:example-phi} to show some polynomial advantage of natively fermionic systems over bosonic ones.
In other words, can we find some useful fermionic algorithm that would necessarily require polynomially deeper local circuits to simulate on a qubit platform? If this is the case, the
present results would make a case in favour of using natively fermionic platforms for addressing fermionic quantum computation and simulation problems.

\begin{acknowledgments}
Discussions with Philippe Faist and Jens Eisert were deeply appreciated. This work has been supported by the German Federal Ministry of Education and Research (BMBF) under the project FermiQP.
\end{acknowledgments}

\bibliographystyle{quantum}
\bibliography{references}

\appendix
\onecolumngrid

\section{Proof of the results} \label{sec:proofs}

\subsection{Proof of Theorem~\ref{thm:no-local-encoding}}
\label{sec:proof-locality}

\begin{theorem1}[Repeated]
A set of local fermionic modes admits a local encoding according to the Definitions~\ref{def:local-encoding} if and only if their locality graph $(V,E)$ is a tree graph.
A tree graph is a connected graph that contains no cycles, that is no sequences of edges $\{(j_1,j_2), (j_2,j_3),\dots,(j_p,j_1)\}$ starting and ending in the same vertex
without any edge appearing more than once.
\end{theorem1}

We are assuming, in our definitions of encodings, that the considered graph is connected. So, to prove the \emph{only if} part, it is sufficient to show that a graph containing
a cycle cannot admit a local encoding. To do this, notice that if the graph contains a cycle, then the operators $\hat{A}_{jk}$ have to satisfy at least one non-trivial relation
of the form~\eqref{eq:group-relations-AB_f}. We will now show that this cannot be satisfied by operators $\hat{A}_{jk}$ which are local according to the tensor product
structure~\eqref{eq:local-hilbert-space} of the Hilbert space (that is, $\hat{A}_{jk}\in\ell_{jk}(\mathcal{H})$) and which anticommute according to condition~\eqref{eq:group-relations-AB_AA}.

Consider indeed the operator $\hat{A}_{jk}$ for any edge $(j,k)\in E$. Due to the locality condition this has to be an operator acting only on $\mathcal{H}_k\otimes\mathcal{H}_j$, 
which can therefore be represented as a tensor with 4 indices, an ingoing and an outgoing index for the Hilbert space $\mathcal{H}_k$ and an ingoing and outgoing index for
the Hilbert space $\mathcal{H}_j$. If we perform a singular value decomposition separating the two indices acting on $\mathcal{H}_k$ from the two indices acting on $\mathcal{H}_j$
we can write
\begin{equation}
    \hat{A}_{jk} = \sum_{i=1}^r \, \lambda_{i}^{(jk)} \: \hat{A}_{i}^{(jk;\,j)} \! \otimes \hat{A}_{i}^{(jk;\,k)} \hspace{15mm} \mbox{for all } (j,k)\in E \,. \label{eq:svd-A} 
\end{equation}
Here, $\lambda_{i}^{(jk)}$ are positive numbers and $r$ is the rank of the singular value decomposition. For every $i=1,\dots,r$, $\hat{A}_{i}^{(jk;\,j)}$ and $\hat{A}_{i}^{(jk;\,k)}$
are operators acting on $\mathcal{H}_j$ and $\mathcal{H}_k$ respectively, which satisfy the orthonormality conditions
\begin{equation}
    \Tr \: {(\hat{A}_{i}^{(jk;\,j)})}^{\!\dag} \hat{A}_{i'}^{(jk;\,j)} = \Tr \: {(\hat{A}_{i}^{(jk;\,k)})}^{\!\dag} \hat{A}_{i'}^{(jk;\,k)} = \delta_{i,i'} \hspace{15mm} \mbox{for all }i,i'=1,\dots,r \label{eq:on-svd-A}
\end{equation} 

Consider now two adjacent edges $(j,k)$ and $(k,l)$. According to relation~\eqref{eq:group-relations-AB_AA} the corresponding operators $\hat{A}_{jk}$ and $\hat{A}_{kl}$
should anticommute. Taking into account the expression~\eqref{eq:svd-A} derived above, this means
\begin{equation}
    0=\{\hat{A}_{jk},\hat{A}_{kl}\}= \sum_{i,i'} \, \lambda_{i}^{(jk)} \lambda_{i'}^{(kl)} \: \hat{A}_{i}^{(jk;\,j)} \! \otimes \{\hat{A}_{i}^{(jk;\,k)},\hat{A}_{i'}^{(kl;\,k)}\} \otimes \hat{A}_{i'}^{(kl;\,l)} \,.
\end{equation}
If we now multiply this expression by $\hat{A}_{i''}^{(jk;\,j)}$ and $\hat{A}_{i'''}^{(kl;\,l)}$ and take the trace over sites $j$ and $l$, due to the orthonormality
conditions~\eqref{eq:on-svd-A} we find $\{\hat{A}_{i}^{(jk;\,k)},\hat{A}_{i'}^{(kl;\,k)}\}=0$ for every $i$, $i'$ and for every pair of
adjacent edges $(j,k)$ and $(k,l)$.

Finally, assuming that the graph contains at least one cycle, let us consider the corresponding operator $i^n A_{j_1,j_2} A_{j_2, j_3} \cdots A_{j_n, j_1}$, which according to
condition~\eqref{eq:group-relations-AB_f} should be equal to the identity. Substituting expression~\eqref{eq:svd-A}, this loop operator can also be written as
\begin{equation}
    i^n \sum_{i_1,\cdots,i_n} \lambda_{i_1}^{(j_1, j_2)} \cdots \lambda_{i_n}^{(j_n, j_1)}\; \left[\hat{A}_{i_1}^{(j_1,j_2;\,j_1)} \hat{A}_{i_n}^{(j_n,j_1;\,j_1)}\right] \: \otimes \: \left[\hat{A}_{i_1}^{(j_1,j_2;\,j_2)},\hat{A}_{i_2}^{(j_2,j_3;\,j_2)}\} \right]
    \: \otimes \dots \otimes \: \left[\hat{A}_{i_{n-1}}^{(j_{n-1},j_n;\,j_n)}  \hat{A}_{i_n}^{(j_n,j_1;\,j_n)} \right] \,. \label{eq:loop-svd}
\end{equation}
This expression however cannot be equal to the identity, as it is actually orthogonal to it with respect to the Hilbert-Schmidt product. The trace of expression~\eqref{eq:loop-svd}
indeed vanishes, because the individual traces of the factors between square brackets vanish. Each factor is the product of two operators which we have shown above to anticommute and,
due to cyclicity, the trace of two anticommuting operators must be equal to zero. 
We conclude that the graph cannot contain any cycle if the operators $\hat{A}_{jk}$ have to satisfy all the relevant conditions of a local encoding according to Definition~\ref{def:local-encoding}.

To prove the \emph{if} part of the theorem, we have to show that it is possible to construct a local encoding for any tree graph. This can be done, for instance, with a version of the
Bravyi-Kitaev encoding~\cite{setia_superfast_2019}. Although this encoding is well know, we will for completeness give here a sketch of the construction, which is quite simple and instructive.

Consider a tree graph $(V,E)$. Let us denote by $m_k$ the degree of each vertex $k$ in the graph, that is the number of edges incident on that vertex. Let us then construct the Hilbert space~\eqref{eq:local-hilbert-space}
by choosing $\mathcal{H}_k= (\mathbb{C}^2)^{\otimes(d_k/2)}$, with $d_k=2\lceil m_k/2 \rceil$. That is, we construct the local Hilbert spaces by attaching to each vertex a qubit
for each pair of incident edges, rounding up by one in case of vertices of odd degree.
On each local Hilbert space $\mathcal{H}_k$ we now define $d_k$ Hermitian operators which all anticommute with each other and which square to $\id$.
We label them as $\gamma_{k,1}, \dots, \gamma_{k,d_k}$. This can be done, for example, by taking 
\begin{align}
    \gamma_{k,i}&=\underbrace{Z\otimes\cdots \otimes Z}_{(i-1)/2 \mbox{ times}} \otimes \, X \otimes \id \otimes \cdots \otimes \id \hspace{2mm}\mbox{ for odd } i \\[3mm]
    \gamma_{k,i}&=\underbrace{Z\otimes\cdots \otimes Z }_{(i-2)/2 \mbox{ times}} \otimes  \, Y \otimes \id \otimes \cdots \otimes \id\hspace{2mm}\mbox{ for even } i \,.
\end{align}
Here, the tensor product refers to the qubits that make up the space $\mathcal{H}_k$ and the operators $X$, $Y$ and $Z$ are the Pauli operators acting on each such qubit.

Let us now introduce an ordering of the edges incident on a given vertex $k$: for every vertex $j$ which is connected to $k$ by an edge of the graph, we assign to $j$ an integer $N_k(j)$,
in such a way that all the vertices directly connected to $k$ will be assigned different integers ranging from $1$ to $m_k$. We do this for all vertices. Finally let us also assign an arbitrary
orientation to each edge: the number $\epsilon_{jk}$ equals $+1$ if $j$ is the tail of the edge $(j,k)$ and $-1$ if $k$ is the tail. We are now ready to construct 
a local encoding into $\mathcal{H}$ of the fermionic modes associated to the graph $(V,E)$. For each edge $(j,k)\in E$ and for each vertex $k\in V$ we define
\begin{align}
    \hat{A}_{jk}&= \epsilon_{jk} \, \gamma_{j,p} \,\gamma_{k,q}   \hspace{25mm}\mbox{with } p=N_j(k)\mbox{ and }  q=N_k(j)\,, \\
    \hat{B}_k&=i^{d_k(d_k-1)/2} \; \gamma_{k,1}\,\gamma_{k,2} \cdots \gamma_{k,d_k} \,.
\end{align}
It is easy to see that these are local Hermitian operators which satisfy relations~\eqref{eq:group-relations-AB_i}--\eqref{eq:group-relations-AB_BB}. As the graph is a tree, there are no
non-trivial cycles, so condition~\eqref{eq:group-relations-AB_f} is also trivially satisfied.

\subsection{Proof of Theorem~\ref{thm:encoding-complexity}}
\label{sec:proof-complexity}

\begin{theorem2}[Repeated]
    Consider the local fermionic modes on a connected graph $(V,E)$ containing an 8-shaped subgraph of size $d$. Consider any local block encoding
    of these modes into a Hilbert space $\mathcal{H}$ according to Definition~\ref{def:gen-local-encoding}. Then the subspace $\mathcal{C}$ does not contain any state that can be produced
    by acting on a product state with a circuit of local two-body gates of depth lower or equal to $d$.
\end{theorem2}

We consider a local block encoding on a graph $(V,E)$ which contains an 8-shaped subgraph of size $d$. Let $\{(i_0,i_1), (i_1,i_2),\dots\}$,
$\{(j_0,j_1), (j_1,j_2),\dots\}$ and $\{(k_0,k_1), (k_1,k_2),\dots\}$ be the sequences of edges that make up this subgraph, as in Definition~\ref{def:8-shaped-subgraph}, where
$i_0\equiv j_0\equiv k_0$ and $i_n\equiv j_m\equiv k_l$. Let $D$ be the integer that satisfies Definition~\ref{def:8-shaped-size} and let $I_D$, $J_D$ and $K_D$ be defined as in equations~\eqref{eq:ID}--\eqref{eq:KD}.

Let us assume now that there exists a state in the subspace $\mathcal{C}$ that can be written as $U\ket{\psi}$, where $\ket{\psi}$ is a product state and $U$ is a circuit of local two-body
unitaries of depth at most $d$. We will show that this leads to a contradiction. Here, the product state condition means that we can write $\ket{\psi}=\bigotimes_{j\in V} \ket{\psi_j}$,
where each $\ket{\psi_j}$ is a normalised local state in $\mathcal{H}_j$. The local two-body nature of the circuit $U$, on the other hand, means that the circuit exhibits a light cone whose
spatial size grows proportionally to the circuit depth. Given that the depth is at most $d$, this means that a local operator, if evolved under $U$, will acquire support only on sites that
are at most at distance $d$ from the support of the original local operator.

We can in particular consider the operators $\widetilde{A}_{jk}=U^\dag\hat{A}_{jk} U$ and $\widetilde{B}_{k}=U^\dag\hat{B}_{k} U$, which will satisfy all
relations~\eqref{eq:gen-relations-AB_i}--\eqref{eq:gen-relations-AB_f} with respect to the product state $\ket{\psi}$, but will have support potentially on all sites within distance $d$ of
the edge $(j,k)$ or the vertex $k$ respectively.
In a similar spirit of enlarging the support of operators by a distance $d$, let us define $\widetilde{I}_D$ as the subset of all vertices of the graph which are within distance $d$ of any 
edge in $I_D$, and similarly for $\widetilde{J}_D$ and $\widetilde{K}_D$. We further define $\widetilde{Q}_D$ as the set of all vertices that belong to more than one of $\widetilde{I}_D$,
$\widetilde{J}_D$ or $\widetilde{K}_D$. 

Consider now the following two operators, corresponding to the loops labelled $L_1$ and $L_2$ in Figure~\ref{fig:figure-8}:
\begin{align}
    L_1&=i^{n+l} \; \widetilde{A}_{i_0i_1} \cdots \widetilde{A}_{i_{n-1}i_n} \, \widetilde{A}_{k_lk_{l-1}} \cdots  \widetilde{A}_{k_1k_{0}}\label{eq:loop1} \\
    L_2&=i^{m+l} \; \widetilde{A}_{j_0j_1} \cdots \widetilde{A}_{j_{m-1}j_m} \, \widetilde{A}_{k_lk_{l-1}} \cdots  \widetilde{A}_{k_1k_{0}} \,. \label{eq:loop2}
\end{align}
As discussed above and considering relation~\eqref{eq:gen-relations-AB_f}, we have $L_1\ket{\psi}=L_2\ket{\psi}=\ket{\psi}$. By taking a partial trace over $\widetilde{Q}_D^c$,
that is over the local Hilbert spaces associated to all vertices of the graph that are not in $\widetilde{Q}_D$, this implies
\begin{align}
    \bigotimes_{j\in\widetilde{Q}_D}\!\ket{\psi_{j}}\! \bra{\psi_{j}}&=\Tr_{\widetilde{Q}_D^c} \ket{\psi}\!\bra{\psi}  \label{eq:tr-loops0}\\
    &=\Tr_{\widetilde{Q}_D^c} L_1\ket{\psi}\!\bra{\psi}L_1^\dag \\
    &=\Tr_{\widetilde{Q}_D^c}   \left(\prod_{(i,i')\in I_D}\!\!\!\!\!\!\widetilde{A}_{ii'}\right) \! \left(\prod_{(k,k')\in K_D}\!\!\!\!\!\!\!\widetilde{A}_{kk'}\right) \ket{\psi}\!\bra{\psi} 
        \left( \prod_{(k,k')\in K_D}\!\!\!\!\!\!\!\widetilde{A}_{kk'}\right)^{\!\!\!\dag} \! \left( \prod_{(i,i')\in I_D}\!\!\!\!\!\!\widetilde{A}_{ii'}\right)^{\!\!\!\dag} \label{eq:tr-loops1} \\
    &=\Tr_{(\widetilde{I}_D \cup \widetilde{K}_D) \cap \widetilde{Q}_D^c}  \! \left(\prod_{(i,i')\in I_D}\!\!\!\!\!\!\widetilde{A}_{ii'}\right) \! \left(\prod_{(k,k')\in K_D}\!\!\!\!\!\!\!\widetilde{A}_{kk'}\right) \!\!\bigotimes_{j\in \widetilde{I}_D \cup \widetilde{K}_D} \!\!\!\! \ket{\psi_j}\!\!\bra{\psi_j}
      \left( \prod_{(k,k')\in K_D}\!\!\!\!\!\!\!\widetilde{A}_{kk'}\right)^{\!\!\!\dag}  \! \left( \prod_{(i,i')\in I_D}\!\!\!\!\!\!\widetilde{A}_{ii'}\right)^{\!\!\!\dag} \!\!. \label{eq:tr-loops2}
\end{align}
Here the product symbols should be understood as a shorthand notation for products of $\widetilde{A}$ operators in the same order as they appear in 
equations~\eqref{eq:loop1} and~\eqref{eq:loop2}.
In step~\eqref{eq:tr-loops1} we have used the cyclicity of the partial trace with respect to operators with support only on $\widetilde{Q}_D^c$ and the 
relations~\eqref{eq:gen-relations-AB_AA} and~\eqref{eq:gen-relations-AB_BB} to eliminate all  operators $\widetilde{A}$ associated to edges not in $I_D$ or $K_D$.
Indeed, by the assumptions of Definition~\ref{def:8-shaped-size}, these edge operators cannot have support in $\widetilde{Q}_D$, as this would imply that there are
vertices in the graph that are within distance $d$ of $R_D$ but also within $\widetilde{Q}_D$ and therefore within distance $d$ of at least two of $I_D$, $J_D$ and $K_D$.
In the last step we have used the product nature of $\ket{\psi}$ and the fact that the remaining operators all have support only in $\widetilde{I}_D \cup \widetilde{K}_D$
to reduce the trace to the sites of $\widetilde{Q}_D^c$ which are also in  $\widetilde{I}_D \cup \widetilde{K}_D$.

Let us now introduce a less cluttered notation to simplify the following analysis. Let us identify four relevant disjoint subsets of vertices of the graph, which we denote as:
\begin{align}
    Q&\equiv\widetilde{Q}_D \\
    A&\equiv\widetilde{I}_D \setminus \widetilde{Q}_D \\
    B&\equiv\widetilde{J}_D \setminus \widetilde{Q}_D \\
    C&\equiv\widetilde{K}_D \setminus \widetilde{Q}_D\,. 
\end{align}
In other words $A$, $B$, $C$ are the vertices which belong to exclusively one of $\widetilde{I}_D$, $\widetilde{J}_D$ or $\widetilde{K}_D$ respectively, and $Q$ are the vertices
which belong to more than one of them. We indicate with $\ket{Q}$ the part of the product state $\ket{\psi}$ within $Q$, that is $\ket{Q}=\bigotimes_{j\in Q}\!\ket{\psi_{j}}$
and analogously for $\ket{A}$, $\ket{B}$, $\ket{C}$. Finally, let us rename the following operators as
\begin{align}
    U_{QA}\equiv\prod_{(i,i')\in I_D}\!\!\!\!\!\!\widetilde{A}_{ii'}\,, \hspace{15mm}
    V_{QB}\equiv\prod_{(j,j')\in J_D}\!\!\!\!\!\!\widetilde{A}_{jj'} \,, \hspace{15mm}
    W_{QC}\equiv\prod_{(k,k')\in K_D}\!\!\!\!\!\!\!\widetilde{A}_{kk'}\,.
\end{align}
Notice that $U_{QA}$, $V_{QB}$ and $W_{QC}$ have support only on $Q\cup A$, $Q\cup B$ and $Q\cup C$ respectively. Notice also that they all anticommute when 
applied to $\ket{\psi}$. This follows from relation~\eqref{eq:gen-relations-AB_AA} observing that they are defined on edges that overlap on a single vertex $i_0\equiv j_0\equiv k_0$.

With this notation, equation~\eqref{eq:tr-loops2} can be expressed as
\begin{equation}
    \ket{Q}\!\bra{Q} = \Tr_{AC} \;   U_{QA} W_{QC} \, \ket{QAC}\!\!\bra{QAC}\, W_{QC}^\dag U_{QA}^\dag \,, 
\end{equation}  
where we simplify the tensor product notation by listing all the states inside the same ket.
This relation implies that the state $U_{QA} W_{QC}\ket{QAC}$ is not entangled with respect to the partition $Q\vert AC$, as the corresponding reduced density matrix is a pure state.
It must thus be a product state. A completely analogous analysis of expression~\eqref{eq:tr-loops0} using $L_2\ket{\psi}=\ket{\psi}$ allows us to conclude the same for
$V_{QB} W_{QC} \ket{QBC}$. That is, in summary, we must have
\begin{align}
    U_{QA} W_{QC}\ket{QAC} &= \ket{Q}\!\ket{\varphi_{AC}} \label{eq:product-UW} \\
    V_{QB} W_{QC}\ket{QBC} &= \ket{Q}\!\ket{\varphi_{BC}} \label{eq:product-VW}
\end{align}
for some suitable normalised states $\ket{\varphi_{AC}}$ and $\ket{\varphi_{BC}}$. The fact that these states are normalised follows from~\eqref{eq:gen-relations-AB_BB}.
We will now show that these last two equations cannot both hold, given the relations~\eqref{eq:gen-relations-AB_AA} and~\eqref{eq:gen-relations-AB_BB}.

These latter relations indeed imply that $U_{QA} V_{QB} W_{QC} \ket{QABC} = - V_{QB} U_{QA} W_{QC} \ket{QABC}$. Substituting~\eqref{eq:product-UW} and~\eqref{eq:product-VW} we would find
\begin{equation}
    \Big(U_{QA} \ket{QA}\!\Big) \ket{\varphi_{BC}} = - \Big(V_{QB} \ket{QB}\!\Big) \ket{\varphi_{AC}}\,. \label{eq:final-anticomm}
\end{equation}
This can only be true if $U_{QA} \ket{QA}$, $V_{QB} \ket{QB}$,  $\ket{\varphi_{BC}}$ and $\ket{\varphi_{AC}}$ are all product states. For instance, taking the partial trace of the last 
equation with respect to the subsystems $AC$ we find that $\Big(\Tr_{A} U_{QA}\ket{QA}\bra{QA}U_{QA}^\dag\Big) \otimes \Big(\Tr_{C} \ket{\varphi_{BC}}\bra{\varphi_{BC}}\Big)$
must be a pure state: this implies that the reduced density matrices of $U_{QA} \ket{QA}$ and $\ket{\varphi_{BC}}$ must both be pure states. The same can be said of $V_{QB} \ket{QB}$
and $\ket{\varphi_{AC}}$ by tracing over $BC$.

So, summarising we have
\begin{align}
    U_{QA} \ket{QA} &= \ket{\theta}\ket{\alpha} \\
    V_{QB} \ket{QB} &= \ket{\overline\theta}\ket{\overline\beta} \\
    \ket{\varphi_{BC}} &= \ket{\beta} \ket{\gamma} \\
    \ket{\varphi_{AC}} &= \ket{\overline\alpha} \ket{\overline\gamma}
\end{align}
for some suitable normalised states $\ket{\alpha}$, $\ket{\overline\alpha}$ living in subsystem $A$, $\ket{\beta}$, $\ket{\overline\beta}$ living in subsystem $B$, $\ket{\gamma}$, $\ket{\overline\gamma}$
living in subsystem $C$ and $\ket{\theta}$, $\ket{\overline\theta}$ living in subsystem $Q$. The fact that they can be chosen to be normalised is again a consequence of~\eqref{eq:gen-relations-AB_BB}.
Substituting this in~\eqref{eq:final-anticomm} we further have $\ket{\theta\,\alpha\,\beta\,\gamma}=-\ket{\overline{\theta}\,\overline{\alpha}\,\overline{\beta}\,\overline{\gamma}}$, which implies 
\begin{equation}
    \ket{\overline\alpha} = \chi_\alpha \ket{\alpha}, \hspace{5mm} \ket{\overline\beta} = \chi_\beta \ket{\beta}, \hspace{5mm} \ket{\overline\gamma} = \chi_\gamma \ket{\gamma}, \hspace{5mm} \ket{\overline\theta} = \chi_\theta \ket{\theta},
\end{equation}
for some phases $\chi_\alpha$, $\chi_\beta$, $\chi_\gamma$, $\chi_\theta$ satisfying  $\chi_\alpha\chi_\beta\chi_\gamma\chi_\theta=-1$.

Consider now again relation~\eqref{eq:product-UW}, rewriting it as $ -W_{QC} U_{QA} \ket{QABC} = \ket{QB}\!\ket{\varphi_{AC}}$. Substituting the results above, this becomes
\begin{equation}
    -W_{QC} \ket{\theta \alpha BC} = \ket{Q \, \overline\alpha \, B \, \overline\gamma} = \chi_\alpha \chi_\gamma\ket{Q\alpha B \gamma}\,, \label{eq:factorised-product-UW}
\end{equation}
which implies $W_{QC} \ket{\theta C}=-\chi_\alpha \chi_\gamma\ket{Q\gamma}$. Similarly, rewriting~\eqref{eq:product-VW} as $ - W_{QC} V_{QB} \ket{QABC} = \ket{QA}\!\ket{\varphi_{BC}}$,
we conclude that 
\begin{equation}
    \ket{QA \beta \gamma} = -W_{QC} \ket{\overline\theta \, A \, \overline\beta \,C} = -\chi_\theta \chi_\beta W_{QC} \ket{\theta \, A \, \beta \,C} \,,\label{eq:factorised-product-VW}
\end{equation}
which implies $W_{QC} \ket{\theta C}=-(\chi_\theta \chi_\beta)^{-1} \ket{Q\gamma}$. Combining these results leads us to $\chi_\alpha \chi_\gamma\ket{Q\gamma} = (\chi_\theta \chi_\beta)^{-1} \ket{Q\gamma}$
which can only be true if $\chi_\alpha\chi_\beta\chi_\gamma\chi_\theta=1$, contradicting our previous conclusion that this product must be negative.

We are therefore forced to abandon our initial assumption, \ie that $U\ket{\psi}\in\mathcal{C}$, which proves the theorem.

\section{Graph with a single cycle} \label{sec:one-loop}

In this section, we provide an example to clarify how the assumptions of Theorem~\ref{thm:encoding-complexity} are strictly necessary. That is, block encodings with a
non-local encoding space $\mathcal{C}$ are really needed only if the graph contains at least two overlapping cycles. Graphs with a single loop indeed admit an encoding
where encoded states can be product states, independently of the loop's size.

\begin{figure*}
\centering
\includegraphics[width=8cm]{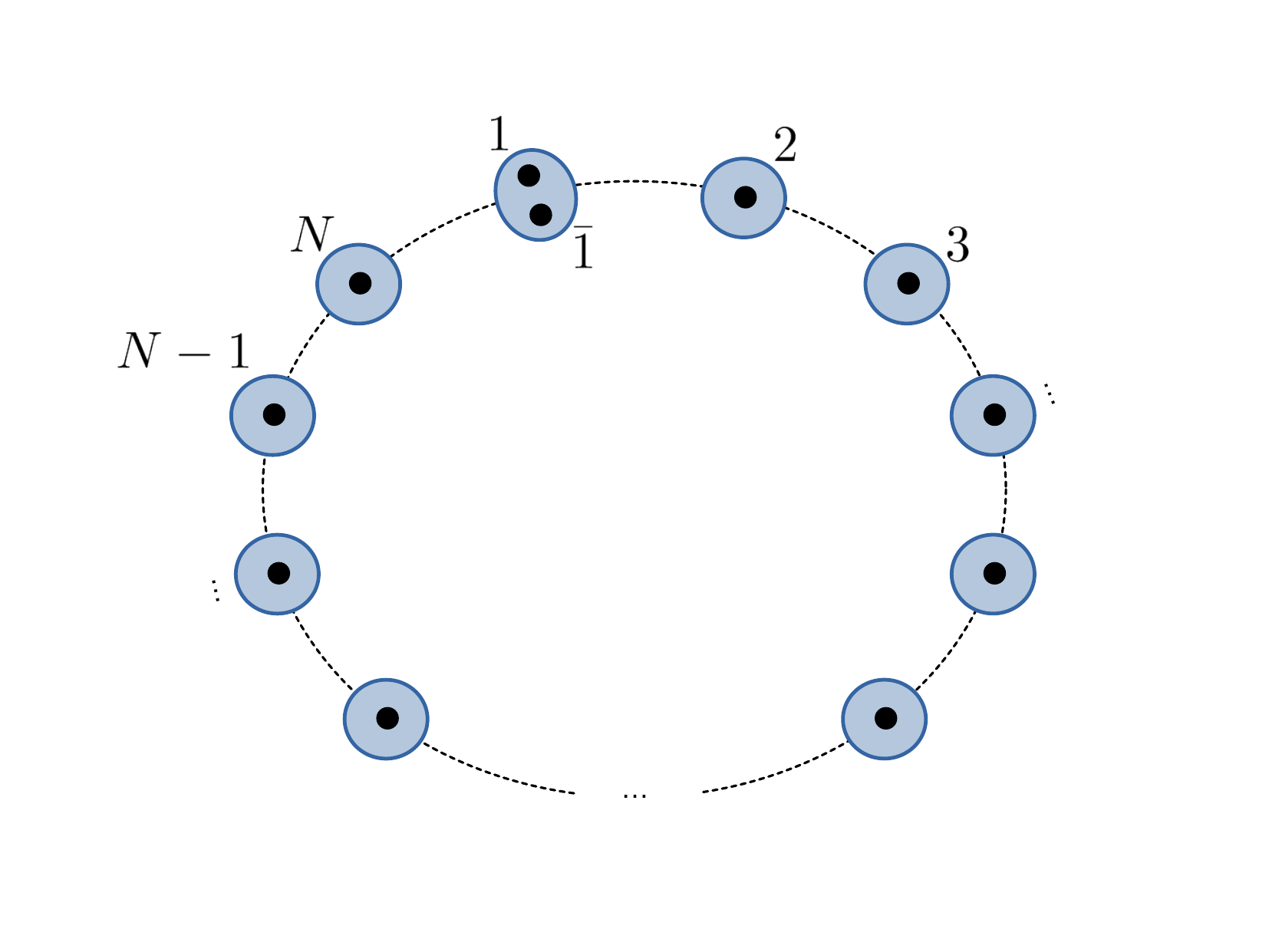}
\caption{A graphical depiction of the system considered in Appendix~\ref{sec:one-loop}. $N$ fermionic modes (blue circles) are positioned on a closed ring. To each mode is associated a 
single qubit (black circles), except for the first mode, to which two qubits are associated.}
\label{fig:ring}
\end{figure*}

Let us show this by constructing an encoding for the simple system of $N$ fermionic modes living on a single ring. That is, consider the graph given by $N$ vertices
labelled $1,\dots,N$ connected in a closed ring geometry, as represented in Figure~\ref{fig:ring}. We associate a fermionic mode to each vertex of the graph. We will now encode 
this fermionic system in the Hilbert space $\mathcal{H}$ constructed in the following way. We associate one qubit to each vertex except for the first vertex. To the first
vertex we associate two qubits, which we label $1$ and $\bar{1}$. So in total $\mathcal{H}$ is composed of $N+1$ qubits, attached to the $N$ vertices of our graph.
We shall denote by $X_k$, $Y_k$, $Z_k$ the Pauli operators acting on the qubit $k$.

Consider now the following local operators associated to each vertex and edge of the graph
\begin{align}
    \hat{B}_k&=Z_k   \hspace{12mm} \forall k=1,\dots,N, \\
    \hat{A}_{k,k+1}&=X_k Y_{k+1} \hspace{5mm} \forall k=1,\dots,N-1, \\
    \hat{A}_{1,N}&=X_N Y_{1} Z_{\bar{1}}\,.
\end{align}
We immediately see that they satisfy relations~\eqref{eq:gen-relations-AB_i}--\eqref{eq:gen-relations-AB_BB} for any state $\ket{\psi}\in\mathcal{H}$.
Relation~\eqref{eq:gen-relations-AB_f} on the other hand is satisfied only on a non-trivial subset of states. In particular equation~\eqref{eq:gen-relations-AB_f}
reduces to
\begin{equation}
    Z_{\bar{1}}Z_{1}Z_{2}\cdots Z_{N}\ket{\psi}=\ket{\psi}\,,
\end{equation}
which is satisfied for any $\ket{\psi}$ in the subspace
\begin{align}
    \mathcal{C}&=\mathrm{span}\left(\left\{\ket{z_{\bar{1}},z_1,\dots,z_n}, \; \forall z\in\{0,1\}^{N+1}\;\mathrm{s.t.}\; \sum_i z_i \: \mathrm{mod}\:2=0 \right\}\right)\,, \label{eq:one-loop-C}
\end{align}
where $\ket{z_{\bar{1}},z_1,\dots,z_n}$ are computational basis states.
We therefore see that the encoding defined by the operators $\hat{A}_{jk}$, $\hat{B}_k$ and the subspace $\mathcal{C}$ above is a local fermionic block encoding
according to Definition~\ref{def:gen-local-encoding}.

Now let us first observe that this encoding is indeed a \emph{block} encoding, as we are encoding the fermionic Fock space into the $2^N$-dimensional subspace
$\mathcal{C}$ of the larger $2^{N+1}$-dimensional Hilbert space $\mathcal{H}$. This is consistent with Theorem~\ref{thm:no-local-encoding}, which forbids a full encoding for
graphs that contain a cycle. Second, it is clear from definition~\eqref{eq:one-loop-C} that the subspace $\mathcal{C}$ contains product states, in fact a whole basis composed of product states. This is compatible with Theorem~\ref{thm:encoding-complexity}, as the graph does not contain any overlapping loops. In particular,
the the fermionic vacuum state $\ket{0}$ is represented by the product state $\ket{0,0,\dots,0}$, which is the $+1$ eigenstate of all the operators $B_k=1-2a^\dag_k a_k$.
It is also possible to rewrite $\mathcal{C}$ as
\begin{align}
    \mathcal{C}&=\mathrm{span}\left(\Big\{\ket{0,z_1,\dots,z_n}, \; \forall z\in\{0,1\}^{N} \;\mathrm{s.t.}\; \sum_{i=1}^N z_i \: \mathrm{mod}\:2=0 \Big\} \right.\nonumber\\
    &\hspace{50mm}  \left.\cup \; \Big\{\ket{1,z_1,\dots,z_n}, \; \forall z\in\{0,1\}^{N} \;\mathrm{s.t.}\; \sum_{i=1}^N z_i \:\mathrm{mod}\: 2=1\Big\} \right) \,.\label{eq:parity-sectors}
\end{align}
This makes it explicit that $\mathcal{C}$ decomposes into two subspaces corresponding to different total fermion number parity,
that is different parities of the observable $\sum_{k=1}^N  B_k$. The role of the additional qubit $\bar{1}$ is essentially to label these two subspaces. 

The construction above can be easily generalised to graphs with multiple loops as long as these loops do not overlap, meaning that the ensuing
conditions~\eqref{eq:gen-relations-AB_f} can be addressed independently.

\end{document}